\documentclass[aps,preprint]{revtex4}
\usepackage{amsmath}
\usepackage{graphicx}
\begin{document}
\title{A DEFINITION FOR FINE TUNING IN ANALOGY TO THE CHAOS}
\author{Zsolt Hetesi}
\affiliation{\small\it  E\"otv\"os  University, Department of Astronomy,
H-1518 Budapest, P.O.Box 32., Hungary }
\email{zs.hetesi@astro.elte.hu},
\author{L\'aszl\'o V\'egh}
\affiliation{\small\it Institute of Nuclear Research of the
Hungarian  Academy of Sciences,
H-4001 Debrecen, Pf. 51, Hungary}
\email{vl@atomki.hu},
\date{\today}
\begin{abstract}
Anthropic principles were grown from the problem of fine tuning.
Although anthropic principles have been discussed in cosmology for years
there is no exact definition for fine tuning.  Starting from the  supposed similarity in the topologies of chaotic and fine tuned regions of the proper phase spaces, we introduce an alternative
Lyapunov indicator for the measure of fine tuning. This fine-tuning indicator
expresses the decrease of life-bearing potentiality of a universe with the
increase of the difference from the physical constants of the universe with
maximum life-bearing potentiality.
\end{abstract}

\maketitle
The Universe and life are connected. Man exists because the Universe is
governed by laws that have led to high levels of organization of matter.
It seems there are apparent coicidences in the
dimensionless basic constants  of the nature. These basic numbers of the nature not
only allow the existence of stable atoms from which matter is built  but also
lead to the formation of galaxies and stars and even more complex stuctures
all are necessary for the existence of life. A small change in the basic
constants would result a universe without these life-supporting conditions.
Therefore we can say that the Universe is finely tuned for life.

Anthropic principles address the question why our Universe has the
fine-tuning property. \cite{hetesi}
Is this a fortunate condition, inevitable or it is expected?
The weak anthropic principle stresses that we, intelligent observers, may
observe only very special properties being compatible with our existence.
The strong anthropic principle claims that the Universe must have those
features which are necessary for life to develop at some stage of its
history. The immediate consequence of the strong anthropic principle  is that
the physical laws and constants must be such as to allow the emergence of life.
There exist several different versions or supplements of strong anthropic
principle. The most known is the design argument which states that life can
occur because of some purposive design. That is the values of physical
constants were selected purposively. For more specific anthropic
definitions see Refs. \cite{barrowtipler, balazs}.

To state that the physical constants of the Universe derive specific values
from an ensemble of different values, we have to suppose the (i) conceptual or
(ii) real existence of the numerical ensemble.
The design argument has chosen the first solution (i), when the other
possibilities exist only as possibilities in the mind of the Designer.
The many-worlds-hypothesis represents the second version (ii), supposing that
an ensemble of other different existing universes is necessary for the
existence of our Universe \cite{barrowtipler}. Both design argument and
many-worlds-hypothesis are created to explain the fine tuning of physical
constants.
There exist some alternative definitions \cite{muller, smolin} but
their common feature is the lack of well-based probability interpretation
for the fine tuning, see the paper of Manson \cite{manson}.

  Life depends on
the presence of the proper chemical elements as building blocks and the
existence of stars which can radiate enough energy for long time for evolution
of  life. If the abundance of carbon and other essential elements are lower
or the number of properly radiating stars are smaller then we have a smaller
probability for the evolution of intelligent life in a universe.
To define fine tuning in a more quantitative way we can study the
probability function for intelligent life of a universe. The  form of
this probability function can be regarded as an anthropic adaptation of the
Drake equation \cite{ellis}. Physical constants are regarded here
as variables.  Therefore
we can present a probability distribution of life-bearing potentiality for
universes  as a function of the basic constans of the physics.

A mathematical expression of fine tuning may stimulate the discussion of anthropic
arguments. Our aim is to construct a simple mathematical definition of fine
tuning analoguous to the Lyapunov-indicator in the chaos theory. (Such a
definition can lead to testable statements therefore arguments of Smolin
\cite{smolin} about the unfalsificability of anthropic principless are
avoidable.)

Discussing the chaotic behavior in the phase space there are regions whith increased sensitivity for the parameter values. This sensitivity in an extreme case is characterized by the Lyapunov indicator $\gamma$ of the given region \cite{Ljapunov}. It
gives a number as result if the motion is chaotic and gives zero if not. Let
us consider two trajectories not far from each other.
The initial distance is $d_0$ and after $t$ the final distance is $d$.
If the distance is groving exponentially, i.e. $d(t) = d_0 exp(\alpha t)$ is
true, then the Lyapunov indicator $\gamma$ is equal to $\alpha$. Generally
the system is chaotic, if in the expression

\begin{equation}
\gamma = \frac {\ln d/d_{0}}{\Delta t} \to \alpha 
\end{equation}

$\gamma > 0$ . If the increase of distance is
smaller, e.g. $d(t) = d_0\beta t$ then $\gamma$ tends to zero.

We suppose that fine tuned region(s) has/have similar topology as chaotic regions. That is there are regions of increased sensitivity in the parameter space describing universes for the life bearing universes and these regions have similar character to a chaotic region. In order to characterize fine tuning we introduce a quantity  like to the
one used in the description of chaotic behavior. 

In the extreme limit of fine tuning, the probability function can be
represented by a Dirac-delta function taken at the point of the parameter
space of physical constants which corresponds to the single life-bearing
universe. The Dirac-delta function $\delta (Q-Q_0)$ can be represented
as the $ n \to \infty $ limit of the Gaussian function

\begin{equation}
\delta (Q-Q_0)=\lim_{n \to \infty} \frac{n}{\sqrt{\pi}} exp\left({-n^2 (Q-Q_0)^2}\right)
\end{equation}

see \cite{diracdelta}.
Now we assume that if there is a fine tuning in the physical parameter $Q$ around the
maximum probability at $Q_0$ then the probability function for the life-bearing
of universes can be approximated by the Gaussian form:

\begin{equation}
p_{\text{life}}(q)= \frac{n}{\sqrt{\pi}} exp(-n^2 q^2)
\label{valseg}
\end{equation}

where $q=\vert Q-Q_0\vert $.
The measure of fine tuning can be defined as

\begin{equation}
\gamma_{\text{fine-tuning}}= -\frac {(\ln p/p_{0})^{1/2}}{\Delta Q} \to n 
\label{fine}
\end{equation}

According to this definition, we have fine tuning, if
$\gamma_{\text{fine-tuning}} > 0$. 
The larger $\gamma_{\text{fine-tuning}}$ the stronger fine-tuning.
Working with dimensionless parameters this definition does not depend 
on the parametrization.

Drawing a paralell between chaos theory and fine tuning is not an arbitrary assumption in this case.
As chaos appears when Lyapunov indicator is not zero fine tuning means that there exist life-bearing
islands in the parameter space of physical constants.

Discussing the production of the carbon and oxigen in the Universe, the
position of the
famous resonance in $C_{12}$ nuclei plays a crucial role. According to
\cite{csoto} outside a narrow window of 0.5 \% and 4 \% of the values of the
strong and nuclear forces, respectively, the stellar production of carbon
or oxigen is reduced by factors of 30 to 1000. These production functions
express Gaussian-like form. We can guess that
the probability to find planets with a proper life-bearing mass would have
even a higher decrease. Now the $\gamma_{\text{fine-tuning}}$ indicator of
Eq. (\ref{fine}) has a definite nonzero value.

\acknowledgments
 The authors are grateful to B\'ela Bal\'azs, Tam\'as Kov\'acs,
 and  Ervin Nemesszeghy for illuminating discussions.
 This work has been supported by the Hungarian OTKA fund No. T37991.

\end{document}